\documentclass[sigconf]{acmart}

\usepackage{booktabs} 
\usepackage{amsfonts}
\usepackage{amssymb}
\usepackage{xspace}
\usepackage{epstopdf}
\usepackage{enumitem}
\usepackage{algorithm}
\usepackage{algorithmic}
\usepackage{multirow}
\usepackage{balance}
\usepackage{booktabs}
\usepackage{array}
\usepackage{wrapfig}
\usepackage{pifont}
\usepackage{pbox}
\usepackage{geometry}
\usepackage{color}

\copyrightyear{2017}
\acmYear{2017}
\setcopyright{acmcopyright}
\acmConference{MM'17}{}{October 23--27, 2017, Mountain View, CA, USA.}
\acmPrice{15.00}
\acmDOI{10.1145/3123266.3123444}
\acmISBN{978-1-4503-4906-2/17/10}

\renewcommand\footnotetextcopyrightpermission[1]{} 

\begin{document}

\def\x{{\mathbf x}}
\def\L{{\cal L}}
\def\eg{\textit{e.g.}}
\def\ie{\textit{i.e.}}
\def\Eg{\textit{E.g.}}
\def\etal{\textit{et al.}}
\def\etc{\textit{etc}}

\title{Affect Recognition in Ads with Application to Computational Advertising}

\author{Abhinav Shukla*}
\thanks{*Indicates equal contribution to this work.\newline}
\affiliation{%
  \institution{International Institute of Information Technology}
  \streetaddress{Gachibowli}
  \city{Hyderabad} 
  \country{India} 
  \postcode{500032}
}
\email{abhinav.shukla@research.iiit.ac.in}

\author{Shruti Shriya Gullapuram*}
\affiliation{%
 \institution{International Institute of Information Technology}
  \streetaddress{Gachibowli}
  \city{Hyderabad} 
  \country{India} 
  \postcode{500032}
}
\email{shruti.gullapuram@students.iiit.ac.in}

\author{Harish Katti}
\affiliation{%
  \institution{Centre for Neuroscience, Indian Institute of Science}
  \streetaddress{Old TIFR building, Gulmohar Marg, Mathikere}
  \city{Bangalore} 
  \country{India}
	\postcode{560012}}
\email{harish2006@gmail.com}

\author{Karthik Yadati}
\affiliation{%
  \institution{Delft University of Technology}
	\streetaddress{Mekelweg 2, 2628 CD}
  \city{Delft}
  \country{Netherlands}
}
\email{n.k.yadati@tudelft.nl}

\author{Mohan Kankanhalli} 
\affiliation{%
 \institution{School of Computing, National University of Singapore}
 \streetaddress{21, Lower Kent Ridge Road}
 \city{Singapore} 
 \postcode{117417}
 \country{Singapore}}
\email{mohan@comp.nus.edu.sg}

\author{Ramanathan Subramanian}
\affiliation{%
  \institution{School of Computing Science, University of Glasgow}
  \streetaddress{43 Woodlands Avenue 9, 08--01}
  \city{Singapore} 
  \country{Singapore}
}
\email{ramanathan.subramanian@glasgow.ac.uk}

%
%
%
\renewcommand{\shortauthors}{A. Shukla et al.}

\begin{abstract}
\textbf{Advertisements} (ads) often include strongly emotional content to leave a lasting impression on the viewer. This work (i) compiles an affective ad dataset capable of evoking coherent emotions across users, as determined from the affective opinions of five experts and 14 annotators; (ii) explores the efficacy of convolutional neural network (CNN) features for encoding emotions, and observes that CNN features outperform low-level audio-visual emotion descriptors~\cite{Hanjalic2005} upon extensive experimentation; and (iii) demonstrates how enhanced affect prediction facilitates computational advertising, and leads to better viewing experience while watching an online video stream embedded with ads based on a study involving 17 users. We model ad emotions based on subjective human opinions as well as objective multimodal features, and show how effectively modeling ad emotions can positively impact a real-life application.  
\end{abstract}

%
%



\maketitle

\vfill\null

\section{Introduction}~\label{sec:intro}

Advertising is a huge and profitable industry and advertisers intend to portray their products or services as not only useful, but also highly desirable and rewarding. Emotions are critical for conveying an effective message to viewers, and have been found to mediate consumer attitudes towards brands~\cite{Holbrook1984,Holbrook1987,Pham2013}. Similar objectives are at play in messages for public health and safety, where certain life choices are portrayed as beneficial and improving one's quality of life, while others are portrayed as harmful and potentially fatal. The ability to objectively quantify advertisements (ads) in terms of emotional content therefore has a wide variety of applications-- \eg, inserting the right type of ads at optimal temporal points within a video stream can beneficially impact both advertisers and consumers in video streaming websites such as YouTube~\cite{cavva,Karthik2013}. Subjective experience of pleasantness (valence) and emotional intensity (arousal) are important affective dimensions~\cite{Russell1980}, and both modulate emotional responses to ads in distinct ways \cite{Broach1995}. Affective content has also been shown to modulate recall of key concepts and episodes in movies \cite{Subramanian2014} and video ads~\cite{cavva}. 

However, affect characterization in ads is a non-trivial problem as with other stimuli such as music and movie clips examined by prior works~\cite{Hanjalic2005,wang2006affective,Koelstra,decaf}. Given that human emotional perception is subjective and the detection of specific emotions such as \textit{joy},\textit{ sorrow} and \textit{disgust} is relatively hard, popular affect recognition (AR) works represent emotions along the valence and arousal dimensions~\cite{Russell1980,greenwald1989}. Also, methods have been devised to estimate emotions in a \textit{content-centric} or \textit{user-centric} manner. \textit{{Content-centric}} methods estimate the emotion evoked by a stimulus by examining audio, visual and textual cues; a popular example is~\cite{Hanjalic2005}. Of late, \textit{{user-centered}} approaches that model the stimulus-evoked emotion via physiological changes induced in the viewer have gained in popularity~\cite{Koelstra,decaf,subramanian2011can}. 
While enabling a fine-grained examination of emotional perception, which is a transient phenomenon, user-centered methods nevertheless suffer from subjectivity limitations.  

This work expressly investigates the modeling of emotions conveyed by ads, and employs subjective human opinions and objective multimedia features to this end. Firstly, upon carefully compiling a diverse set of 100 ads, we examine the efficacy of this ad dataset to coherently evoke emotions across viewers. To this end, we compare the affective opinions of five experts and 14 novice annotators
, and find that the two groups are highly concordant. Secondly, we explore the utility of Convolutional Neural Networks (CNNs) for encoding audio-visual emotional features. As the compiled ad dataset is relatively small and insufficient for CNN training, we employ \textit{domain adaptation} to transfer affective knowledge gained from the LIRIS-ACCEDE movie dataset~\cite{baveye2015liris} for modeling ad emotions. Extensive experimentation confirms that the synthesized CNN descriptors outperform popular audio-visual features proposed in~\cite{Hanjalic2005} especially for valence recognition. Thirdly, we show how accurate encoding of the ad emotions can facilitate optimized insertion of ads into streaming video, used for income generation by online websites such as YouTube. A user study with 17 viewers confirms that the insertion of emotionally relevant ads within the streamed video can maximize viewer experience. In summary, we make the following research contributions:

\begin{itemize}[noitemsep,nolistsep]
\item[1.] Ours is one of the few works to examine AR in ads, and the only work to characterize ad emotions in terms of subjective human opinions and objective audio-visual features. 
\item[2.] We explore the utility of CNNs for encoding ad emotions. We show the effectiveness of a new CNN, \textbf{\textit{AdAffectNet}} (AAN) generated by fine-tuning the \textit{Places205} CNN architecture~\cite{places14} for AR. For fine-tuning and domain adaptation, we have employed the extensively annotated LIRIS-ACCEDE movie dataset~\cite{baveye2015liris}. Extensive experiments reveal that the AAN features outperform emotional audio-visual descriptors proposed in~\cite{Hanjalic2005}, and the best AR performance is achieved with multi-task learning which exploits audio-visual similarities among emotionally homogeneous ads.  
\item[3.] We show how improved affect modeling with the CNN features can facilitate the CAVVA ad-in-video insertion strategy~\cite{cavva}. Our AAN model can especially predict ad valence better than the baseline~\cite{Hanjalic2005}, which positively impacts viewer experience.
\end{itemize} 

From here on, Sec~\ref{RW} discusses related work, while Sec~\ref{ad_set} presents the compiled ad dataset and related statistics. Section~\ref{CM} describes our AAN model and associated AR experiments. Sec~\ref{US} presents a user study to evaluate an ad-insertion strategy~\cite{cavva} with different affect encoding methods, while Sec~\ref{CFW} concludes the paper.

\section{Related Work}\label{RW}
To highlight the novelty of our work, we briefly review prior works examining (i) AR, and (ii) the impact of ad-evoked emotions. 

\subsection{Affect Recognition}
Many approaches have been devised to infer the emotions evoked by multimedia stimuli in a \textit{content-centric} or \textit{user-centric} manner. Content-centric approaches~\cite{Hanjalic2005,wang2006affective} predict the elicited emotion by examining audio-visual cues in the analyzed stimuli. In contrast, user-centric AR methods~\cite{Koelstra,decaf,subramanian2016ascertain} predict the stimulus-evoked emotion by measuring physiological changes in users (or content consumers). Nevertheless, both content and user centric methods require labels identifying stimulus emotion, and these labels are compiled from reliable annotators whose affective opinions are generally \textit{acceptable}, given human subjectivity in emotion perception. 

\subsection{Emotional impact of ads}
A number of works have studied the impact of ad-induced emotions on user behavior~\cite{Holbrook1984,Holbrook1987,Pham2013}. Holbrook and Batra~\cite{Holbrook1987} remark that emotions induced by ads strongly modulate users' brand attitude. Pham \etal~\cite{Pham2013} conclude that ad-evoked feelings impact users both explicitly and implicitly. Ad-evoked emotions are found to \textit{change} user attitude towards (especially hedonistic) products.   

While a body of works have examined the correlation between ad emotions and user behavior, very few works have exploited these findings for developing targeted advertising mechanisms. The only work that incorporates emotional information for modeling context in advertising is CAVVA~\cite{cavva}, where arousal and valence evoked by video scenes are estimated via~\cite{Hanjalic2005} to identify optimal ads and corresponding insertion points which maximize user engagement.

\subsection{Analysis of related work}
Examination of the literature reveals that (1) AR studies are hampered by the subjectivity in emotion perception, and a control dataset that can coherently evoke emotions across users is necessary for effectively learning content or physiology-based emotion predictors; (2) Despite the well-known impact of ad emotions on user behavior, there has hardly been any attempt to incorporate emotion-related findings in a computational advertising framework.  

In this regard, we present the first work to compile a control set of affective ads, which elicit concordant affective opinions from experts and naive users. Also, we synthesize CNN-based emotion descriptors which are found to outperform audio visual features proposed in~\cite{Hanjalic2005}. We also perform a user study and show how better affect encoding can facilitate the ad-insertion framework~\cite{cavva} to improve viewing experience. Details pertaining to our ad dataset are presented below.  



\section{Advertisement Dataset}\label{ad_set}
\begin{figure}[t]
\includegraphics[width=0.32\linewidth]{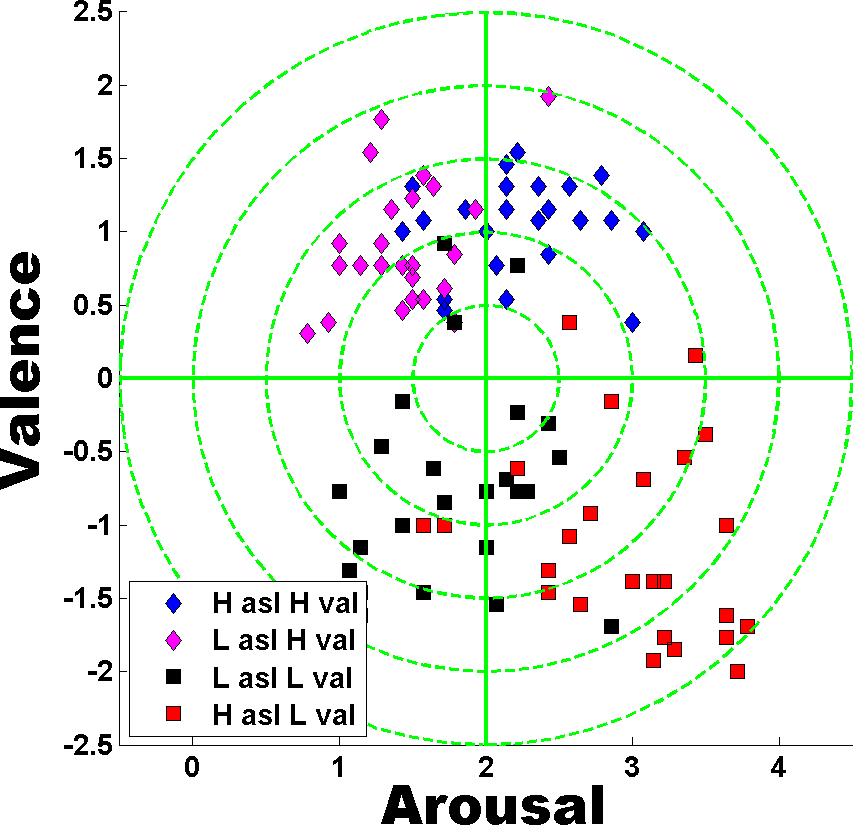}\hfill
\includegraphics[width=0.32\linewidth]{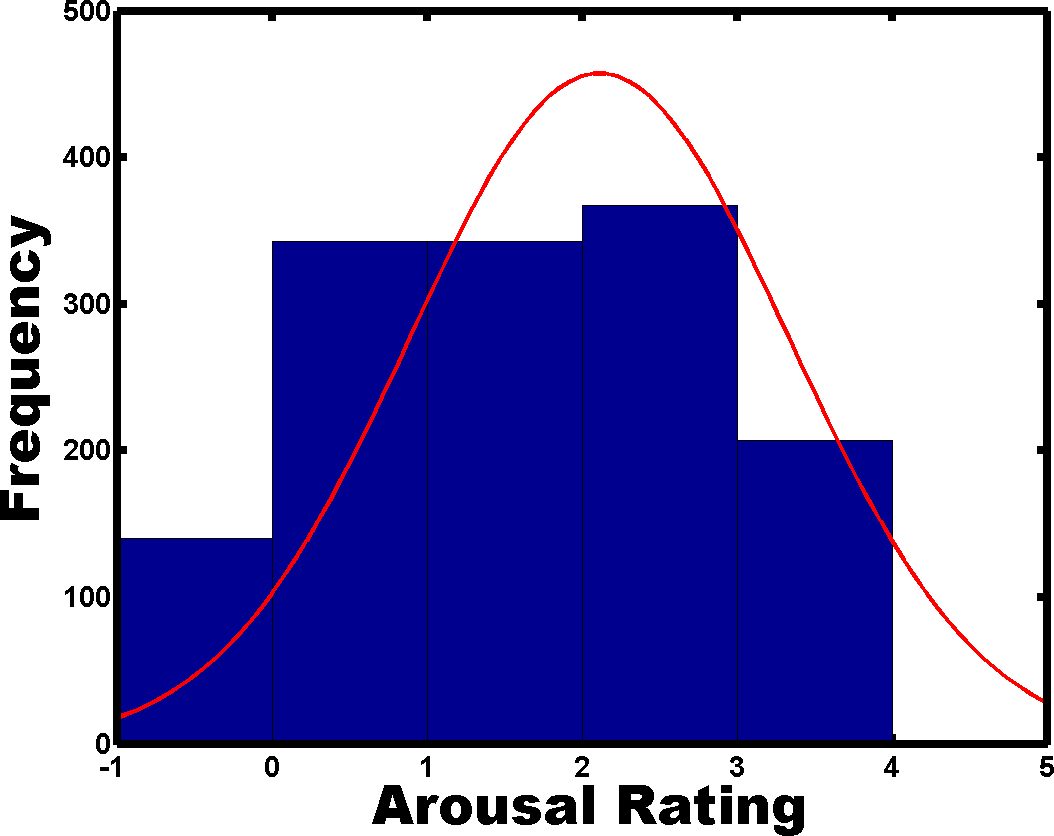}\hfill
\includegraphics[width=0.32\linewidth]{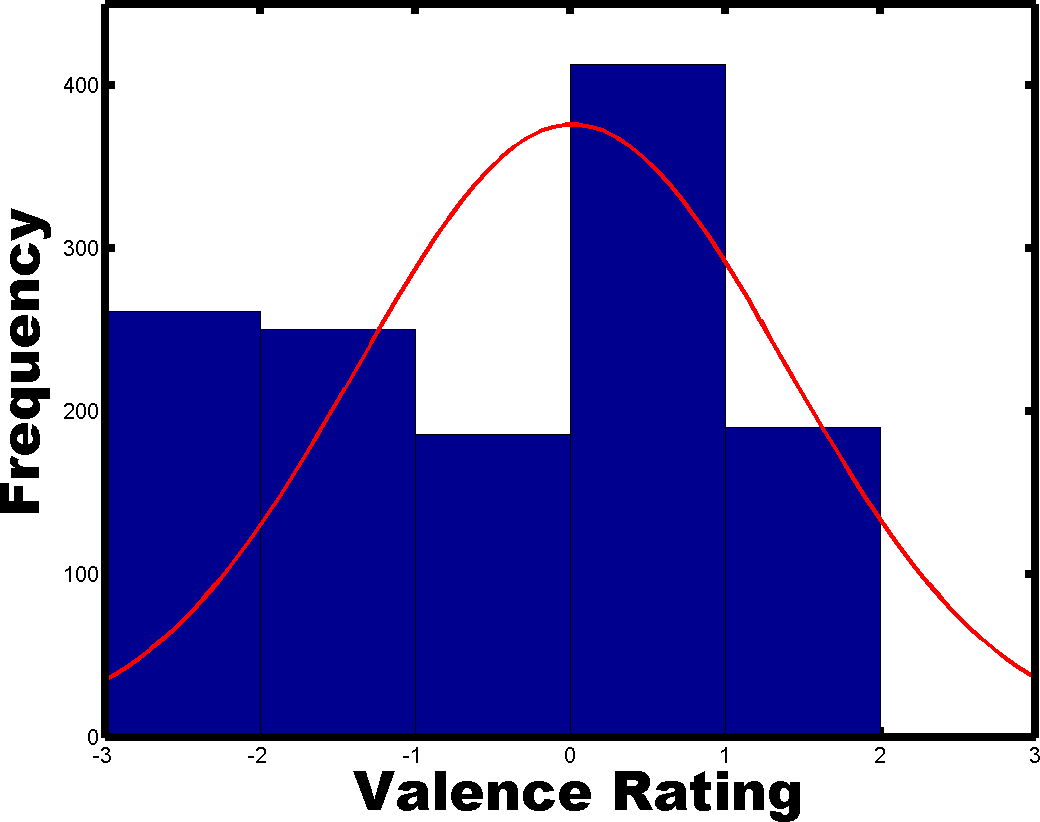}\vspace{-.1cm}
\caption{\label{Annot_dist} (left) Scatter plot of mean asl, val ratings color-coded with expert labels. (middle) Asl and (right) Val rating distribution with Gaussian pdf overlay (view under zoom).}\vspace{-.5cm}
\end{figure}

Defining \textbf{\textit{valence}} as the feeling of \textit{pleasantness}/\textit{unpleasantness}, \textbf{\textit{arousal}} as the \textit{intensity of emotional feeling} and \textbf{\textit{engagement}} as the \textit{level of interest} while viewing an audio-visual stimulus, five experts carefully compiled a dataset of 100, roughly 1-minute long commercial advertisements (ads) which are used in this work. These ads are publicly available on online video websites and found to be uniformly distributed over the arousal--valence plane defined by Greenwald \textit{et al.}~\cite{greenwald1989} (Fig.~\ref{Annot_dist}). An ad was chosen if there was consensus among all five experts on its valence and arousal labels (defined as either \textit{high} (H)/\textit{low} (L)). The high valence ads typically involved product promotions, while low valence ads were social messages depicting the ill effects of smoking, alcohol and drug abuse, \etc. Labels provided by experts were considered as \textbf{\textit{ground-truth}}, and used for all recognition experiments in this work. 
 
To evaluate the effectiveness of the ads as control stimuli, \ie, examine how consistently they evoke emotions across viewers, 14 novice users rated the ads for valence (val), arousal (asl) and engagement (eng) upon familiarization. All ads were rated on a 5-point scale, which ranged from -2 (\textit{very unpleasant}) to 2 (\textit{very pleasant}) for val, 0 (\textit{calm}) to 4 (\textit{highly aroused}) for asl and 0 (\textit{boring}) to 4 (\textit{highly engaging}) for eng. Table~\ref{tab:ads_des} presents summary statistics for ads over the four quadrants. Evidently, low val ads are longer and are perceived as more arousing than high val ads suggesting that they evoked stronger emotional feelings among viewers.

\begin{table}[t]
\vspace{-.1cm}
\fontsize{7}{7}\selectfont
\renewcommand{\arraystretch}{1.6}
\caption{\label{tab:ads_des} Summary statistics for quadrant-wise ads.}\vspace{-.2cm}
\centering
\begin{tabular}{|c|cccc|} \hline
\textbf{Quadrant} & \textbf{Mean length (s)} & \textbf{Mean asl} & \textbf{Mean val} &  \textbf{Mean eng} \\ \hline \hline
\textbf{H asl, H val} & 48.16 & 2.17 & \ 1.02 & 2.50\\
\textbf{L asl, H val} & 44.18 & 1.37 & \ 0.91 & 2.23\\
\textbf{L asl, L val} & 60.24 & 1.76 & -0.76 & 2.47\\
\textbf{H asl, L val} & 64.16 & 3.01 & -1.16 & 2.56\\ \hline
\end{tabular}
\vspace{-.3cm}
 \end{table}
 
We also computed inter-rater agreement in terms of the (i) Krippendorff's $\alpha$ and (ii) Cohen's $\kappa$ measures. The $\alpha$ coefficient is applicable when multiple raters code data with ordinal scores-- we obtained $\alpha = 0.60, 0.37$ and $0.23$ for val, asl and eng implying valence impressions were most consistent across raters. We then computed the $\kappa$ agreement between annotator and ground-truth labels to determine concordance between the novice and expert groups. To this end, we thresholded each rater's asl, val scores by their mean rating to assign H/L labels for each ad, and compared these labels with the ground truth. This procedure revealed a mean agreement of 0.84 for val and 0.67 for asl across raters. We also computed $\kappa$ between the annotator and expert \textit{populations} by thresholding the mean asl, val score per ad across raters against the grand mean-- this method returned $\kappa = 0.94$ for val and 0.67 for asl\footnote{Chance agreement corresponds to a $\kappa$ value of 0.}. Clearly, there is good-to-excellent agreement between annotators and experts on affective impressions with considerably higher concordance for val. The observed concordance between the independent expert and annotator groups affirms that the compiled 100 ads are suitable control stimuli for affective studies.  

Another desirable property of a control affective dataset is the independence of the asl and val dimensions. To this end, we (i) examined scatter plots of the annotator ratings, and (ii) computed correlations amongst those ratings. The scatter plot of the mean asl, val annotator ratings, and the distribution of asl and val ratings are presented in Fig.~\ref{Annot_dist}. The scatter plot is color-coded based on expert labels, and interestingly is different from the classical `C' shape observed with images~\cite{IAPS}, music videos~\cite{Koelstra} and movie clips~\cite{decaf} owing to the difficulty in evoking medium asl/val but strong val/asl responses. The distributions of asl and val ratings are also roughly uniform resulting in Gaussian fits with large variance, with modes observed at median scale values of 2 and 0 respectively. A close examination of the scatter plot reveals that a number of ads are rated as moderate asl, but high/low val. This can be attributed to the fact that ads are constrained to effectively convey a positive or negative message to viewers, which is not typically true of images or movie scenes. Finally, Wilcoxon rank sum tests on annotator ratings revealed significantly different asl ratings for high and low asl ads ($p<0.00005$), and distinctive val scores for high and low valence ads ($p<0.000001$), consistent with expectation.    

Pearson correlations among the asl, val and eng dimensions from the annotator ratings as shown in Table~\ref{Corr_Ratings}. As each rater assesses multiple ads, we corrected for multiple comparisons by limiting the false discovery rate to within 5\% via the procedure described in~\cite{benjamini1995controlling}. Bold values in Table~\ref{Corr_Ratings} denote correlations found to be significant ($p<0.05$) over all raters. Rating correlations reveal that the asl and val dimensions are only weakly correlated, while a positive and significant correlation is noted between asl and eng in line with prior studies~\cite{Koelstra,decaf}. Analyses from here on will focus on the asl and val attributes given the strong connection between asl and eng. The user study (Sec~\ref{US}) will focus on user engagement/experience. 

Overall, examination of ad labels reveals that (i) Our ads constitute a control dataset for affective studies as asl and val ratings are largely uncorrelated; (ii) Also, different from the `C'-shape characterizing the asl-val relationship for other stimulus types, asl and val ratings are  uniformly distributed for the ad stimuli, and (iii) There is considerable concordance between the expert and annotator populations on affective labels, implying that the selected ads effectively evoke coherent emotions across viewers. The next section describes content-centric ad AR.

\begin{table}[htbp]
\vspace{-.1cm}
\fontsize{8}{8}\selectfont
\renewcommand{\arraystretch}{1.2}
\caption{\label{Corr_Ratings} Mean correlations between self-rated attributes. Significant correlations ($p<0.05$) are denoted in bold.} \vspace{-.1in}
\centering
\begin{tabular}{l|lll}
                     & \textbf{asl} & \textbf{val} & \textbf{eng} \\ \hline \hline
\textbf{Arousal}     &  1    & -0.19    & \textbf{0.36}     \\
\textbf{Valence}     & ~    & ~1         & 0.11       \\
\textbf{Engagement}  & ~    & ~          & ~1         \\
\hline
\end{tabular}
\vspace{-.2in}
\end{table}

\section{Computational Model}\label{CM}

This section describes the datasets and the computational model employed for recognizing emotions conveyed by our ads. 

\subsection{Datasets}
Due to subjective variance in emotion perception, careful affective labeling is imperative for effectively learning content-centric~\cite{Hanjalic2005,wang2006affective} or user-centric~\cite{Koelstra,decaf} affective correlates, which is why we analyze ads that evoked perfect consensus among experts. Of late, convolutional neural networks (CNNs) have become extremely popular for visual~\cite{alex12} and audio~\cite{Huang2014} recognition, but these models require huge training data. Given the small size of our ad dataset, we fine-tune the pre-trained \textit{Places205}~\cite{places14} model using the affective LIRIS-ACCEDE movie dataset~\cite{baveye2015liris}, and employ this fine-tuned model for encoding ad emotions-- a process known as \textit{\textbf{domain adaptation}} in machine learning literature.

\begin{sloppypar}
To learn deep features for modeling ad affect, we employed the \textit{Places205} CNN~\cite{places14} intended for image classification. \textit{Places205} is trained using the {Places-205} dataset comprising 2.5 million images and 205 scene categories. The {Places-205} dataset contains a wide variety of scenes with varying illumination, viewpoint and field of view, and we hypothesized a strong relationship between scene perspective, lighting and the scene mood. \textbf{LIRIS-ACCEDE} contains asl, val ratings for $\approx$ 10 s long movie snippets, whereas our ads are about a minute-long (ranging from 30--120 s).
\end{sloppypar}

\subsection{FC7 Feature Extraction via CNNs}
For ad AR, we represent the \textit{visual} modality using \textit{key-frame} images, and the \textit{audio} modality using \textit{spectrograms}. We fine-tune \textit{Places205} via the LIRIS-ACCEDE~\cite{baveye2015liris} dataset to synthesize \textbf{\textit{AdAffectNet}} (AAN), and use the fully connected layer (fc7) AAN descriptors for our analysis.

\subsubsection{Keyframes as Visual Descriptors}
From each video in the ad and LIRIS-ACCEDE datasets, we sample one \textit{key frame} every three seconds-- this enables extraction of a continuous video profile for AR. This process generated 1791 key-frames for our 100 ads.

\subsubsection{Spectrograms as Audio Descriptors}
\begin{figure}[t]
\centering
\includegraphics[width=2.8cm]{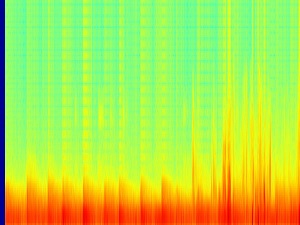}\hfill
\includegraphics[width=2.8cm]{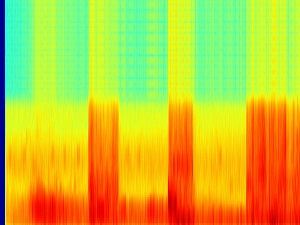}\hfill
\includegraphics[width=2.8cm]{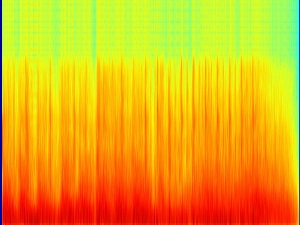}
\centerline{\textbf{L asl, H val} \hspace{1.2cm} \textbf{H asl, L val} \hspace{1.2cm} \textbf{H asl, H val}}\vspace{-.1cm}
\caption{Exemplar spectrograms for varied emotional ads. $x$ denotes time (0-10 s), while $y$ denotes spectral magnitude observed at each time instant. High and low frequency densities are respectively shown in red and green shades.}
\label{fig:spects}
\vspace{-.3cm}
\end{figure}

Spectrograms (SGs) are visual representations of the audio frequency spectrum, and have been successfully employed for AR from speech and music~\cite{baveyethesis}. Specifically, transforming the audio content to a spectrogram image allows for audio classification to be treated as a visual recognition problem. We extract spectrograms over the 10 s long LIRIS-ACCEDE clips, and consistently from 10 s ad segments. This process generates 610 spectrograms for our ad dataset. Following~\cite{baveyethesis}, we combine multiple tracks to obtain a single spectrogram (as opposed to two for stereo). Each spectrogram is generated using a 40 ms window short time Fourier transform (STFT), with 20 ms overlap. Fig.~\ref{fig:spects} shows three exemplar spectrograms indicative of emotional ad content. Note greater densities of high frequencies in high asl ads, and such intense scenes are often characterized by sharp frequency changes. 

\subsubsection{CNN Training}
We use the Caffe~\cite{caffe} deep learning framework for fine-tuning \textit{Places205}, with a momentum of 0.9, weight decay of 0.0005, and a base learning rate of 0.0001 reduced by $\frac{1}{10}^{th}$ every 20000 iterations. We totally train four binary classification AAN networks this way to recognize high and low asl/val from audio/visual features. To fine-tune \textit{Places205}, we use only the top and bottom 1/3rd LIRIS-ACCEDE videos in terms of asl and val rankings under the assumption that descriptors learned for the extreme-rated clips will effectively represent affective concepts. 4096-dimensional \textit{fc7} outputs extracted from the four AAN networks are used for ad AR. 

\subsection{AR with audio-visual features}
We will mainly compare our AAN network based AR framework against the algorithm of Hanjalic and Xu~\cite{Hanjalic2005} in this work. Even after a decade since it was proposed, this algorithm remains one of the most popular AR baselines as noted from recent works such as~\cite{Koelstra,decaf}. In ~\cite{Hanjalic2005}, asl and val are modeled via low-level descriptors describing motion activity, colorfulness, shot change frequency, voice pitch and sound energy in the scene. These hand-crafted features are interpretable, and employed to estimate time-continuous asl and val levels conveyed by the scene. Table~\ref{tab:exp_det} summarizes the audio-visual features used in our AR experiments, and the proportion of positive audio/video frames for val and asl in our ad dataset.

%
%
%

\begin{table}[t]
\fontsize{6}{6}\selectfont
\renewcommand{\tabcolsep}{4.6pt}
\caption{Extracted features and +ve class proportions (in \%) for the audio and visual modalities.} \label{tab:exp_det} \vspace{-.2cm} 
\begin{center}                                                         
\begin{tabular}{@{}|c|ccc|@{}} 

\hline
\textbf{Attribute} & \multicolumn{3}{c|}{\textbf{Valence/Arousal}} \\ 

~ &{\textbf{Audio}}& {\textbf{Video}}&{\textbf{aud+vid (A+V)}}\\ \hline\hline
 
 {\textbf{AAN}} & 4096D AAN FC7   & 4096D AAN FC7 features by  & 8192D AAN FC7 features \\

 \textbf{Features} & features obtained  & extracted from keyframes  &  with SGs + keyframes \\ 
        &  with 10s SGs.   & sampled every 3 seconds.            & over 10s intervals. \\ \hline

\textbf{Hanjalic~\cite{Hanjalic2005}} & Per-second sound  & Per-second shot change & Concatenation of \\ 
\textbf{Features} &  energy and pitch  & frequency and motion & audio-visual features. \\
 & statistics~\cite{Hanjalic2005}. & statistics~\cite{Hanjalic2005}. & ~ \\ \hline

\textbf{val/asl} & {43.8}/{51.9} & {43.4}/{51.6} & {43.8}/{51.9}\\ 
\textbf{+ve class prop (\%)} &  &  & \\ \hline
\end{tabular}
\end{center}
\vspace{-.3cm}
\end{table}


\subsection{Experiments and Results}
We first provide a brief description of the classifiers used and settings employed for our binary AR experiments, where the objective is to assign  a binary (H/L) label for asl and val evoked by each ad, using the extracted fc7/low-level audio visual features. Experimental results will be discussed thereafter.

\paragraph*{Classifiers:} We employed the Linear Discriminant Analysis (LDA), linear SVM (LSVM) and Radial Basis SVM (RSVM) classifiers for AR. LDA and LSVM attempt to separate H/L labeled training data with a hyperplane, while RSVM is a non-linear classifier which separates H and L classes, linearly inseparable in the input space, via transformation onto a high-dimensional feature space.

In addition to the above \textit{single-task learning} methods which do not exploit the underlying structure of the input data, we also explored the use of \textit{multi-task learning} (MTL) for AR. When posed with the learning of multiple \textit{related} tasks, MTL seeks to jointly learn a set of task-specific classifiers on modeling task relationships, which is highly beneficial when learning with few examples. Among the MTL methods available as part of the MALSAR package~\cite{zhou2012mutal}, we employed the sparse graph-regularized MTL (SR-MTL) where \textit{a-priori} knowledge regarding task-relatedness is modeled in the form of a graph $R$. Given tasks $t = 1...T$, with $X_t$ denoting training data for task $t$ and $Y_t$ their labels, SR-MTL jointly learns a weight matrix $W = [W_1..W_T]$ such that the objective function $\sum_{t=1}^{T} \Vert W_t^T X_t -Y_t\Vert_F^2+\alpha \Vert WR \Vert^2_{F}+ \beta \Vert W \Vert_1+\gamma \Vert W \Vert^2_F$ is minimized. Here, $\alpha, \beta, \gamma$ are regularization parameters, while  $\Vert.\Vert_F$ and $\Vert.\Vert_1$ denote the matrix Frobenius ($\ell_2$) and $\ell_1$-norms respectively.

\begin{figure*}[t]
\centerline{\includegraphics[width=0.33\linewidth,height=3cm]{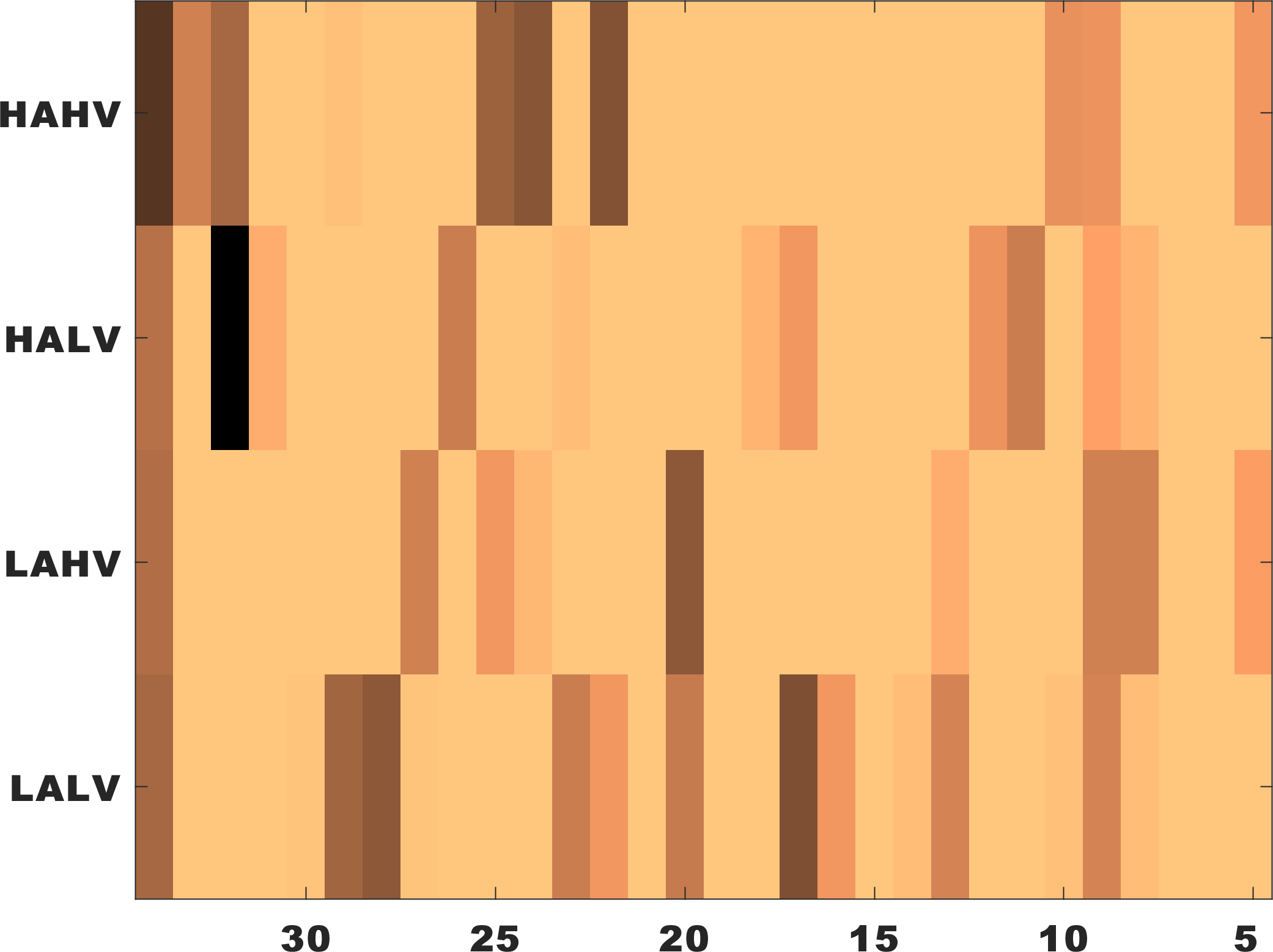}\hspace{0.05cm}
\includegraphics[width=0.33\linewidth,height=3cm]{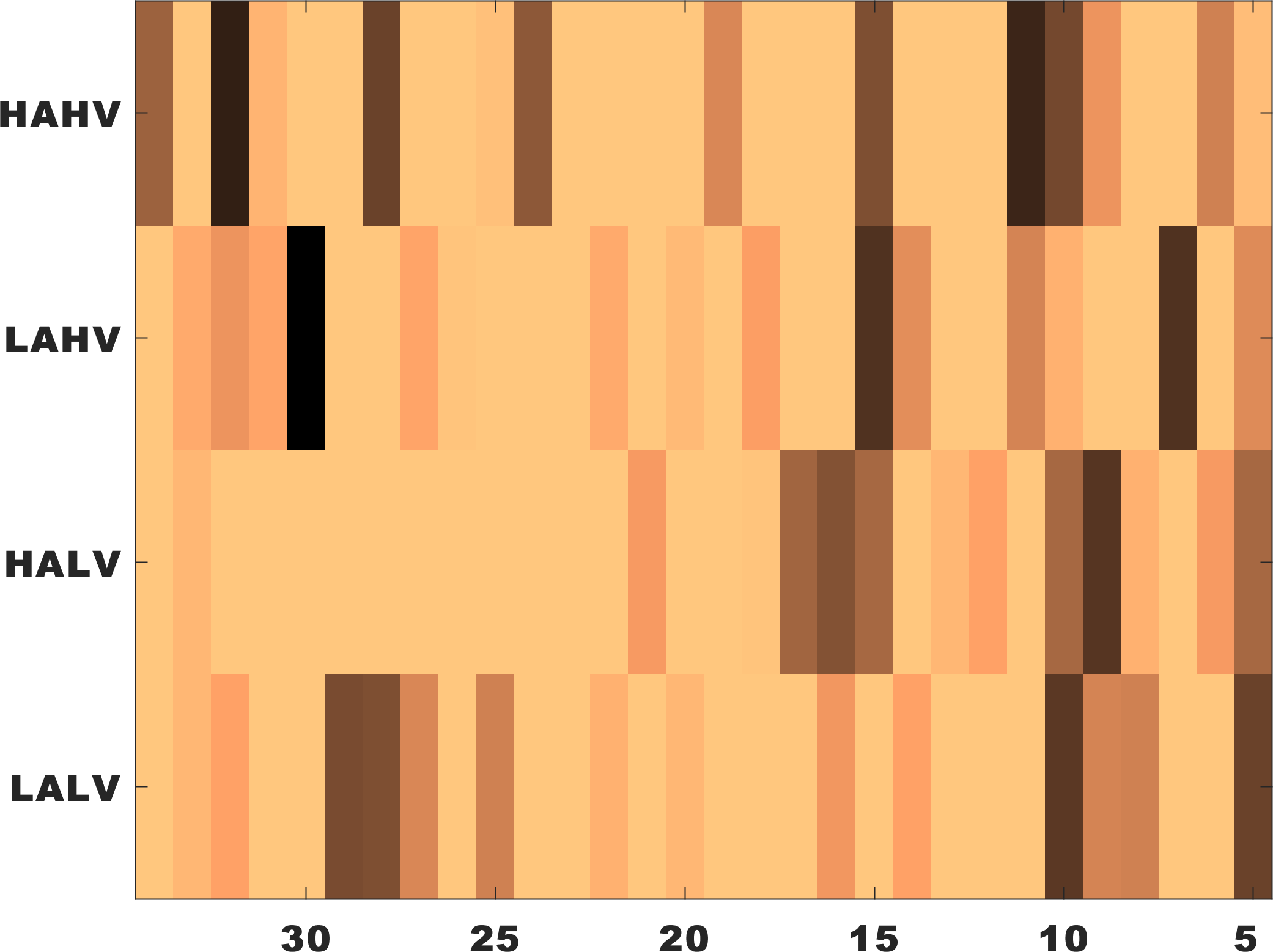}\hspace{0.05cm}
\includegraphics[width=0.33\linewidth,height=3cm]{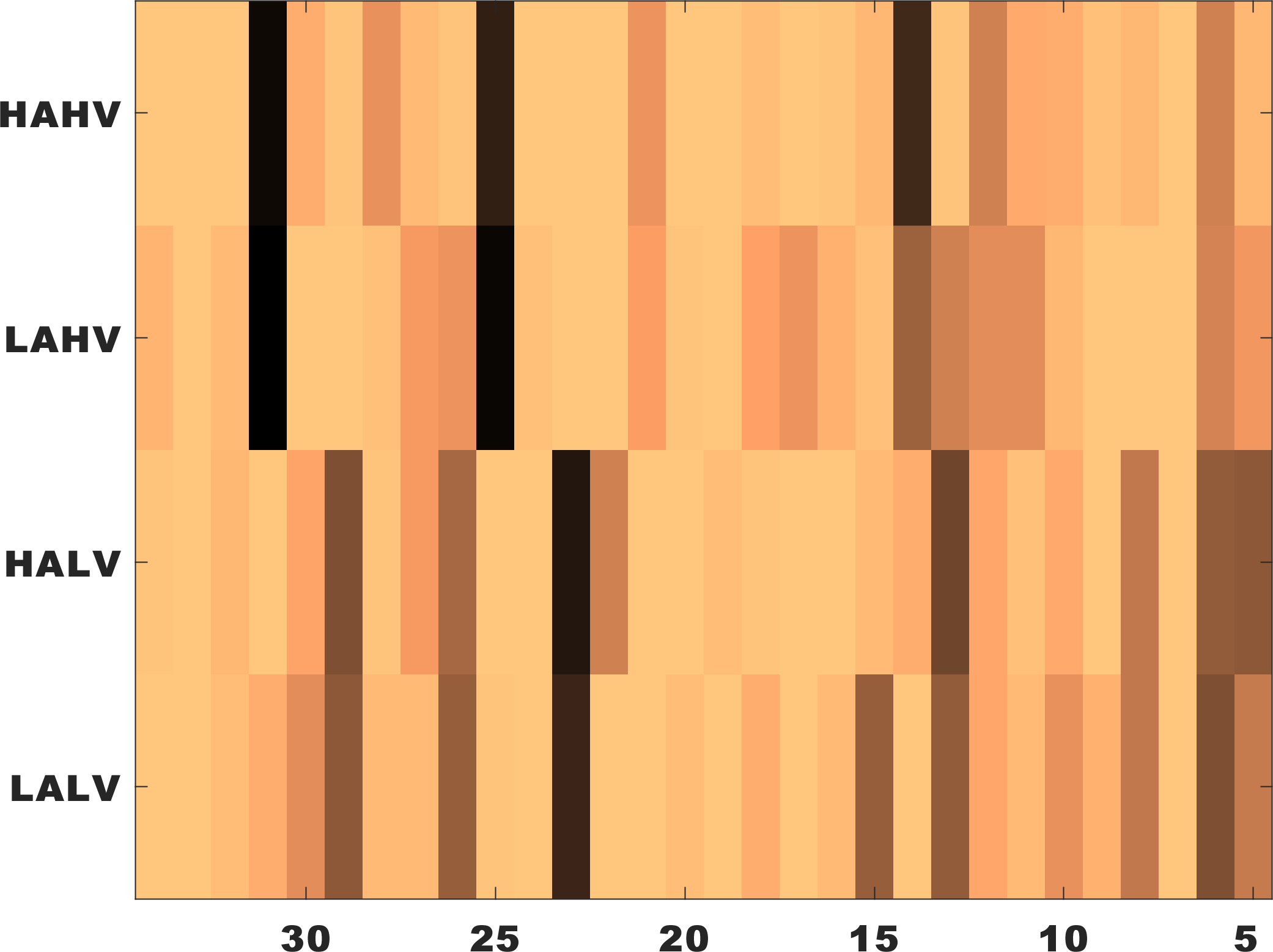}}
\centerline{\textbf{Shot Frequency}\hspace{0.21\linewidth}\textbf{Pitch Amplitude}\hspace{0.23\linewidth}\textbf{Motion Activity}} \vspace{-.2cm} 
\caption{\label{MTL_ex} Learned MTL weights for  the four quadrants (tasks) when fed with the specified low-level features computed over the final 30 s of the 100 ads.}
\vspace{-.2cm}
\end{figure*}

MTL is particularly suited for dimensional AR, and one can expect similarities in terms of audio-visual content among high val or high asl ads. We exploit underlying similarities by modeling each asl-val quadrant as a task (\ie, all H asl, H val ads will have identical task labels). Also, quadrants with same asl/val labels are deemed as related tasks, while those with dissimilar labels are considered unrelated. The graph $R$ guides the learning of $W_t$'s, as shown in the three examples in Fig.\ref{MTL_ex}, where SR-MTL is fed with the specified features computed over the final 30 s of all ads. Darker shades denote salient MTL weights. Shot change frequency is found to be a key predictor of asl in~\cite{Hanjalic2005}, and one can notice salient weights for H asl H val ads in particular. The attributable reason is that our H asl H val ads involve frequent shot changes to maintain emotional intensity, while the mood of our H asl L val ads\footnote{which depict topics like drug and alcohol abuse, and overspeeding.} is strongly influenced by semantics. Likewise, pitch amplitude is deemed as a key val predictor, and salient weights can be consistently seen over the 30 s temporal window for HV ads. Finally, more salient weights for H val ads with the motion activity feature implies that our positive val ads involve accentuated motion.   

\paragraph*{Metrics and Experimental Settings:} We used the F1-score (F1), defined as the harmonic mean of precision and recall as our performance metric, given our unbalanced dataset (Table~\ref{tab:exp_det}). Apart from unimodal (audio (A) or visual (V)) fc7 features, we also employed feature fusion and probabilistic decision fusion of the unimodal outputs. Feature fusion (A+V) involved concatenation of fc7 A and V features over 10 s windows (see Table~\ref{tab:exp_det}), while the $W_{est}$ technique~\cite{koelstra2012fusion} is employed for decision fusion (DF). In DF, the test label is computed as $\sum_{i=1}^2 \alpha_i^*t_ip_i$, where $i$ indexes the A,V modalities, $p_i$'s denote posterior A,V classifier probabilities and $\{\alpha_i^*\}$ are the optimal weights maximizing test F1-score, and determined via a 2D grid search. If $F_i$ denotes the training F1-score for the $i^{th}$ modality, then $t_i = \alpha_i F_i/\sum_{i=1}^2 \alpha_i F_i$ for given $\alpha_i$.

As the Hanjalic (Han) algorithm~\cite{Hanjalic2005} uses audio plus visual features to model asl and val, we only consider (feature and decision) fusion performance in this case. As we evaluate AR performance on a small dataset, we present AR results over 10 repetitions of 5-fold cross validation (CV). CV is typically used to overcome the \textit{overfitting} problem on small datasets, and the optimal classifier parameters (including regularization parameters for MTL) are determined from the range $[10^{-3},10^{3}]$ via an inner five-fold CV on the training set. Finally, in order to examine the temporal variance in AR performance, we present F1-scores obtained over (a) all ad frames (`All'), (b) last three frames (L3) and (c) last frame (L)\footnote{This equates to estimating the ad asl/val typically over the terminal 30/10 s, when one would expect the conveyed emotion to be strongest.}.     

\begin{table*}[t]
\fontsize{8}{8}\selectfont
\renewcommand{\arraystretch}{1.3}
\centering
\caption{Ad AR from content analysis. F1 scores are presented in the form $\mu \pm \sigma$. } \label{tab:cap} \vspace{-.2cm}
\begin{tabular}{|c|ccc|ccc|}
  \hline
	\multicolumn{1}{|c|}{\textbf{Method}} & \multicolumn{3}{c|}{\textbf{Valence}} & \multicolumn{3}{c|}{\textbf{Arousal}} \\ \hline 
	\multicolumn{1}{|c|}{~} & {\textbf{F1 (all)}} & {\textbf{F1 (L3)}} & {\textbf{F1 (L)}}  & {\textbf{F1 (all)}} & {\textbf{F1 (L3)}} & {\textbf{F1 (L)}}\\ \hline

	
	\textbf{Audio FC7 + LDA}  & 0.61$\pm$0.04 & 0.62$\pm$0.10 & 0.55$\pm$0.18 & 0.65$\pm$0.04 & 0.59$\pm$0.10 & {0.53$\pm$0.19}\\
	\textbf{Audio FC7 + LSVM} & 0.60$\pm$0.04 & 0.60$\pm$0.09 & 0.55$\pm$0.19 & 0.63$\pm$0.04 & 0.57$\pm$0.09 & 0.50$\pm$0.18\\
	\textbf{Audio FC7 + RSVM} & {0.64$\pm$0.04} & \textbf{0.66$\pm$0.08} & {0.62$\pm$0.17} & \textbf{0.68$\pm$0.04} & {0.60$\pm$0.10} & {0.53$\pm$0.19}\\ \hline

	\textbf{Video FC7 + LDA} & 0.69$\pm$0.02 & 0.79$\pm$0.08 & {0.77$\pm$0.13} & 0.63$\pm$0.03 & 0.58$\pm$0.10 & 0.57$\pm$0.18\\
	\textbf{Video FC7 + LSVM}& 0.69$\pm$0.02 & 0.74$\pm$0.08 & 0.70$\pm$0.15 & 0.62$\pm$0.02 & 0.57$\pm$0.09 & 0.52$\pm$0.17\\
	\textbf{Video FC7 + RSVM}& {0.72$\pm$0.02} & \textbf{0.79$\pm$0.07} & 0.74$\pm$0.15 & \textbf{0.67$\pm$0.02} & {0.62$\pm$0.10} & {0.58$\pm$0.19}\\ \hline
	
	\textbf{Audio FC7 + MTL} & 0.85$\pm$0.02 & 0.83$\pm$0.10 & 0.78$\pm$0.20& 0.78$\pm$0.03 & 0.62$\pm$0.14 & 0.45$\pm$0.16\\ 
	\textbf{Video FC7 + MTL} &\textbf{0.96$\pm$0.01} & 0.94$\pm$0.07 & 0.82$\pm$0.25 & \textbf{0.94$\pm$0.01} & 0.87$\pm$0.12 & 0.63$\pm$0.29 \\ \hline

	\textbf{{A+V FC7 + LDA}} &  0.70$\pm$0.04 & 0.66$\pm$0.08 & 0.49$\pm$0.18 & 0.60$\pm$0.04 & 0.52$\pm$0.10 & {0.51$\pm$0.18}\\
	\textbf{A+V FC7 + LSVM}  &  0.71$\pm$0.04 & 0.66$\pm$0.07 & 0.49$\pm$0.19 & 0.56$\pm$0.04 & 0.49$\pm$0.10 & 0.47$\pm$0.19\\
	\textbf{A+V FC7 + RSVM}  &  \textbf{0.75$\pm$0.04} & {0.70$\pm$0.07} & {0.55$\pm$0.17} & \textbf{0.63$\pm$0.04} & {0.56$\pm$0.11} & 0.49$\pm$0.19\\ \hline
	
\textbf{{A+V Han + LDA}} & 0.59$\pm$0.09 &{0.63$\pm$0.08} & {0.64$\pm$0.12} & {0.54$\pm$0.09} & 0.50$\pm$0.10 & {0.58$\pm$0.08}\\
	\textbf{A+V Han + LSVM}  & {0.62$\pm$0.09} & {0.62$\pm$0.10} & {0.65$\pm$0.11} & 0.55$\pm$0.10 & {0.51$\pm$0.11} & 0.57$\pm$0.09\\
	\textbf{A+V Han + RSVM}  & \textbf{0.65$\pm$0.09} & {0.62$\pm$0.11} & 0.62$\pm$0.12 & \textbf{0.59$\pm$0.12} & {0.58$\pm$0.11} & 0.56$\pm$0.10\\ \hline
		
	\textbf{{A+V FC7 LDA DF}} & 0.60$\pm$0.04  &  {0.66$\pm$0.04}  &  {0.70$\pm$0.19} & 0.59$\pm$0.02 & 0.60$\pm$0.07 & {0.57$\pm$0.15} \\
	\textbf{A+V FC7 LSVM DF}  & {0.65$\pm$0.02}  &  {0.66$\pm$0.04}  &  0.65$\pm$0.08 & {0.60$\pm$0.04}  & {0.63$\pm$0.10} & {0.53$\pm$0.13} \\
	\textbf{A+V FC7 RSVM DF}  & \textbf{{0.72$\pm$0.04}}  &  {0.70$\pm$0.04}  &  {0.70$\pm$0.12} & {0.69$\pm$0.06} & \textbf{0.75$\pm$0.07} & {0.70$\pm$0.07}\\ \hline
	
\textbf{{A+V Han LDA DF}}   & 0.58$\pm$0.09 & 0.58$\pm$0.09 & \textbf{0.61$\pm$0.09} & {0.59$\pm$0.06} & {0.59$\pm$0.07} & {0.61$\pm$0.08}\\
	\textbf{A+V Han LSVM DF}  & 0.59$\pm$0.10 & 0.59$\pm$0.09 & 0.60$\pm$0.10 & {\textbf{0.61$\pm$0.05}} & {0.61$\pm$0.08} & {0.60$\pm$0.09}\\
	\textbf{A+V Han RSVM DF}  & 0.60$\pm$0.08 & 0.56$\pm$0.10 & {0.58$\pm$0.09} & {0.58$\pm$0.09} & {0.56$\pm$0.06} & {0.58$\pm$0.09}\\ \hline

	\textbf{A+V FC7 + MTL}  & \textbf{{0.89$\pm$0.03}} & {0.88$\pm$0.11} & {0.77$\pm$0.26} & \textbf{{0.87$\pm$0.03}} & {0.68$\pm$0.17} & {0.46$\pm$0.20} \\
	%
	
	\textbf{A+V Han + MTL}  & 0.77$\pm$0.04 & 0.79$\pm$0.07 & 0.74$\pm$0.15 & 0.78$\pm$0.04 & 0.73$\pm$0.11 & 0.58$\pm$0.22\\ \hline
  

  \hline
\end{tabular}
\vspace{-.2cm}
\end{table*}

\subsubsection{Results Overview}
Table~\ref{tab:cap} presents the asl, val F1-scores under the various settings. The highest F1 over all the considered temporal windows, achieved with the single-task and multi-task classifiers, and via feature/decision fusion are denoted in bold. Based on the observed results, we make the following claims. 

Focusing on \textbf{\textit{unimodal fc7 features and single-task classifiers}}, val (peak F1 = 0.79) is generally recognized better than asl (peak F1 = 0.68) and especially with video features. A and V fc7 features perform comparably for asl. Much higher asl, val recognition scores are achievable with the \textit{\textbf{MTL classifier}} (F1 of 0.96 for val and 0.94 for asl) due to its ability to exploit the underlying similarities among audio and visual features among similarly labeled ads. MTL F1-scores are consistently higher with V features for both asl and val. 

Concerning recognition with \textbf{\textit{single-task classifiers}} and \textbf{\textit{fused fc7 features}}, comparable or better F1 scores are achieved with multimodal approaches. In general, better recognition is achieved via decision fusion as compared to feature fusion\footnote{The relative efficacy of feature or decision fusion depends on the specific problem and features on hand.}. For val, the best fusion performance (0.75 with feature fusion and RSVM classifier) is superior compared to A-based (F1 = 0.66), but inferior compared to V-based (F1 = 0.79) recognition. Contrastingly for asl, fusion F1-score (0.75 with DF) considerably outperforms unimodal methods (0.68 with A, and 0.67 with V). Focusing on the \textbf{\textit{MTL classifer}}, MTL F1-scores in the \textbf{A+V FC7 + MTL} condition are considerably higher than single-task F1-scores\footnote{We are not aware of MTL-based decision fusion methods.} analogous to the unimodal case, even though the best F1-scores are still less than those achieved with video fc7 features $+$  MTL. 

Comparing A$+$V \textit{\textbf{fc7 vs Han}} features, fc7 descriptors clearly outperform Han features with both single and multi-task approaches. The difference in performance is prominent for val, while comparable recognition is achieved with both features for asl. RSVM produces the best F1-scores for both asl and val among \textbf{\textit{single-task classifiers}} with unimodal and multimodal approaches. However, the linear \textit{\textbf{MTL}} model considerably outperforms all single-task methods with both fc7 and Han features. These observations suggest that while the H and L asl/val features for \textit{all ads} are difficult to linearly classify \textit{per se}, exploiting underlying similarities among \textit{quadrant-specific ads} enables better linear separability.   

Relatively small $\sigma$ values are observed for the `All' condition with the adopted five-fold CV procedure in Table~\ref{tab:cap}, especially with fc7 features suggesting that the AR results are \textit{\textbf{minimally impacted}} by \textit{\textbf{ overfitting}}. Examining \textit{\textbf{temporal windows}} considered for AR, higher $\sigma$'s are observed for the L3 and L cases, which denote model performance on the terminal ad frames. Surprisingly, one can note a general degradation in asl recognition for the L3 and L conditions with A/V features, while val F1-scores are more consistent. Also, a sharp degradation in performance is noted with MTL for the L3 and L conditions. Three inferences can be made from the above observations, namely, (1) Greater heterogeneity in the ad content towards endings is highlighted by the large variance with fusion and MTL-based approaches; (2) Fusion models synthesized with Han features appear to be more prone to overfitting, given the generally larger $\sigma$ values seen with the corresponding models; (3) That asl recognition is lower in the L3 and L conditions highlights the limitation of using a \textit{single} asl/val label (as opposed to dynamic labeling) over time. Generally lower F1-scores achieved for asl with all methods suggests that asl is a more transient phenomenon as compared to val, and that coherency between val features and labels is sustainable over time.           

\subsubsection{Discussion}
As ads are inherently emotional and have great influencing/monetizing capacity~\cite{Holbrook1987,Pham2013}, the ability to infer ad emotions and make optimal ad insertions within video streams would be highly advantageous for multimedia systems. Therefore, it is surprising that very few works~\cite{videosense,cavva} have attempted to mine visual and emotional content in ads. In this regard, our work expressly sets out to model the emotion conveyed by 100 ads based on subjective human opinions and objective audio-visual features. 

We carefully curated a small but diverse set of ads based on consensus among experts, and examined if those ads could coherently evoke emotions across viewers by acquiring asl and val ratings from 14 annotators. A good-to-excellent agreement on asl and val impressions is noted between the expert and annotator groups. Also, annotator ad ratings are found to be  uniformly distributed over the asl-val plane, with only a weak negative correlation noted between asl and val ratings.   

As the compiled ads are found to constitute a control affective dataset, we modeled the emotion conveyed by these ads in terms of audio-visual features. Specifically, we extracted fc7 layer outputs from the \textit{AdAffectNet} CNNs fed with key frames and spectrograms for video and audio-based affect modeling. While CNNs have been previously used for video and audio-based AR~\cite{baveye2015liris,Fan2016}, the modeled scenes are only a few second long snippets. In contrast, we have explored the validity of CNN-based AR for full-length ads, some of which are over a minute long, in this work. 

Obtained AR results confirm that the synthesized fc7 features are effective predictors of asl and val. They outperform the audio-visual features proposed by Hanjalic and Xu~\cite{Hanjalic2005} with both single and multi-task classifiers. In particular, while fc7 features are considerably better for val, Han features provide competitive performance for asl. Optimal AR is achieved with the MTL classifier, which is able to effectively exploit the underlying similarities among emotionally homogeneous ads in terms of audio visual content. Nevertheless, a significant drop in recognition performance is generally noted for the terminal ad portion with most methods, and especially for asl, implying that (a) asl is a more transient phenomenon as compared to val, and there is less coherence between the employed AV features and asl labels towards the ad endings, and (b) the use of a single asl/val over the entire ad duration may be inappropriate, and one may need to acquire time-varying labels for affective studies. The next section will evaluate whether the superior AR achieved with AAN fc7 features translates to optimized ad insertion in a computational advertising task via a user study.

\section{Computational Advertising- User Study}\label{US}

Given the superior AR achieved by our \textit{AdAffectNet} CNN features, we hypothesized that this should in turn enable optimized selection and insertion of affective ads within streamed video content, as discussed in the CAVVA ad insertion framework~\cite{cavva}. Video-in-video advertising is complex, as it aims to strike a balance between (a) maximizing ad impact, and (b) minimally disrupting (or ideally, enhancing) viewing experience while watching
a program video into which the ads are embedded. Also, while ad insertion strategies~ have modeled ad-video relevance in terms of low-level visual context~\cite{videosense} and high-level emotional context~\cite{cavva}, their performance has not been compared against human context assessment. 

To this end, we performed a user study to evaluate whether the ad insertion framework formulated in~\cite{cavva}, which employs the affect estimation methodology of Hanjalic and Xu~\cite{Hanjalic2005}, would benefit from better affect prediction via our deep AAN features. Better estimation of the affect induced by the video content and candidate ads can enable optimized selection of ads and corresponding insertion points. Specifically, we compared video program sequences generated via the CAVVA framework by estimating arousal (asl) and valence (val) scores for the ads and video scenes via (a) the baseline method of Hanjalic and Xu~\cite{Hanjalic2005}, (b) our deep AAN model and (c) human annotators. 

\subsection{Dataset}\label{US-DS} 

For the user study, we chose 28 ads (from the 100 used in this work) and three program videos. The program videos were scenes from a television sitcom \textit{Friends} (\textit{\textbf{friends}}) and two movies \textit{The Pursuit of Happyness} (\textit{\textbf{ipoh}}) and \textit{Children of Heaven} (\textit{\textbf{coh}}), with predominantly social themes and situations invoking high-to-low val and asl. Summary statistics of the three program videos are presented in Table \ref{tab:progdetails}. Each program video was segmented into 8 scenes, and the average scene length was 118 seconds. We obtained val and asl scores for the video scenes and 28 ads using (a) normalized softmax class probabilities~\cite{bishop:2013} output by our AAN model, with video and audio fc7 features respectively used for val and asl estimation (b) the baseline model (Han)~\cite{Hanjalic2005} and (c) ratings from three experts (Manual). We then inserted ads into each program video based on method-specific affect scores  with the optimization strategy described in~\cite{cavva}, and obtained 9 unique \textit{\textbf{video program}} sequences (mean length 19.4 min) comprising the inserted ads. Exactly 5 ads were inserted in each program video, and 21 of the 28 chosen ads were cumulatively inserted at least once onto the 9 video programs (upon being selected via any of the three methods), with an average insertion frequency of 2.14.

\begin{table}[t]
\fontsize{8}{8}\selectfont
\renewcommand{\arraystretch}{1.3}
\centering
\caption{Summary of program video statistics.}\vspace{-.2cm}
\begin{tabular}{|c|c|cc|}
  \hline
	\multicolumn{1}{|c|}{\textbf{Name}} & \multicolumn{1}{c|}{\textbf{Scene length (s)}} & \multicolumn{2}{c|}{\textbf{Manual Rating}} \\
	 \hline 
	\multicolumn{1}{|c|}{~} & {~} & {\textbf{Valence}} & {\textbf{Arousal}}\\ \hline
	\textbf{coh}  & 127$\pm$46 & {0.08$\pm$1.18} & {1.53$\pm$0.58}\\
	\textbf{ipoh}  & 110$\pm$44 & {0.03$\pm$1.04} & {1.97$\pm$0.49}\\
	\textbf{friends}  & 119$\pm$69 & {1.08$\pm$0.37} & {2.15$\pm$0.65}\\
\hline
\end{tabular}
\vspace{-.2cm}
\label{tab:progdetails}
\end{table}

\subsection{Experiment and Questionnaire Design}~\label{US-Ques}
We recruited a total of 17 users (5 female, mean age 20.5 years) to evaluate the video program sequences. Each user saw one exemplar sequence corresponding to the three affect prediction strategies. We followed a randomized $3 \times 3$ Latin square design so that all nine video programs were covered with three users.

Our user evaluation was in two parts; In line with the twin goals underlying seamless ad insertion within streaming video, we evaluated whether the ad insertion strategy resulted in  (a) increased brand recall, and (b) minimal disturbance and improved viewing experience. We performed recall evaluation by measuring the impact of the ad insertion strategy on \textit{immediate} and \textit{day-after} recall. These \textit{objective} measures quantified the impact of ad insertion on short-term (immediate) and long-term (day-after) memory of viewers, on viewing the video programs. Specifically, we measured the  proportion of (i) inserted ads that were recalled correctly (\textit{Correct} recall), (ii) inserted ads that were not recalled (\textit{Forgotten}) and (iii) non-inserted ads incorrectly recalled as viewed, perhaps owing to their inherent salience (\textit{Incorrect} recall). For those ads that were inserted into program sequences and were correctly recalled, we also assessed whether viewers perceived them to be contextually appropriate with respect to program content.

The viewer was provided with a key-frame visual from each of the 28 ads, as well as a response sheet for every video program sequence. In addition to the recall related questions, we asked viewers to indicate whether they perceived the correctly recalled ads as being inserted at an appropriate position in the video stream (\textit{Good insertion})\footnote{This was one way of inferring if the ad placements facilitated their recall.}. All recall and insertion quality-related responses were acquired from users as binary values. We pooled responses from viewers after they had watched video sequences generated via deep AAN, Han and manual affective scores for analyses.  

While increased ad recall reflects a key desired effect of a successful ad insertion strategy, ads that are out of sync with the video program flow may disrupt viewer experience. In some cases, this disruptiveness may indirectly contribute to the recall, but would adversely impact viewing experience. So, mere recall alone does not indicate optimal ad insertion, relevance of the ad to the program or an enhanced viewer experience. To address these issues, we defined a second set of \textit{subjective} experience evaluation measures and asked users to provide ratings on a Likert scale of 0--4 for the following questions, with 4 implying \textit{best} and 0 denoting \textit{worst}:

\begin{itemize}[noitemsep,nolistsep]
\item[1.]	Whether advertisements were uniformly distributed across the video program?
\item[2.]	Whether the inserted ads blended well with the flow of the video program?
\item[3.]	Whether the inserted ads had a content and mood similar to surrounding program?
\item[4.]	What was the overall viewer experience while each video program?
\end{itemize}

Each participant filled the recall and experience-related questionnaires (provided in supplementary material) after watching each video program. They also filled in the day-after  recall questionnaire, a day after completing the experiment.

\subsection{Results and Discussion}

\begin{figure*}
\centerline{\includegraphics[width=0.33\linewidth,height=3cm]{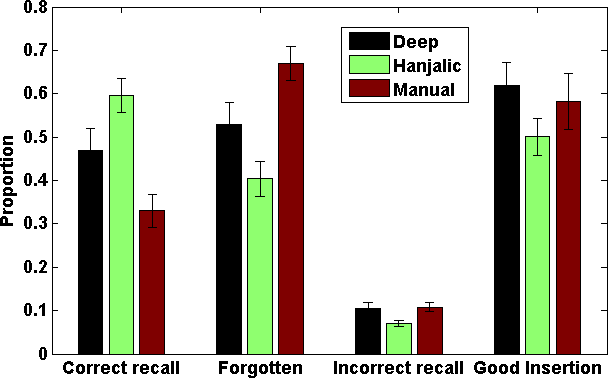}\hspace{0.05cm}
\includegraphics[width=0.33\linewidth,height=3cm]{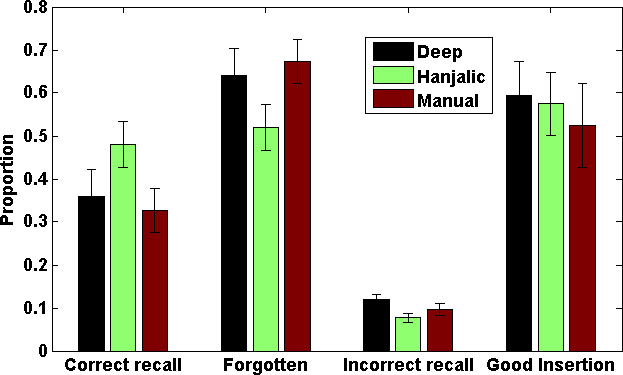}\hspace{0.05cm}
\includegraphics[width=0.33\linewidth,height=3cm]{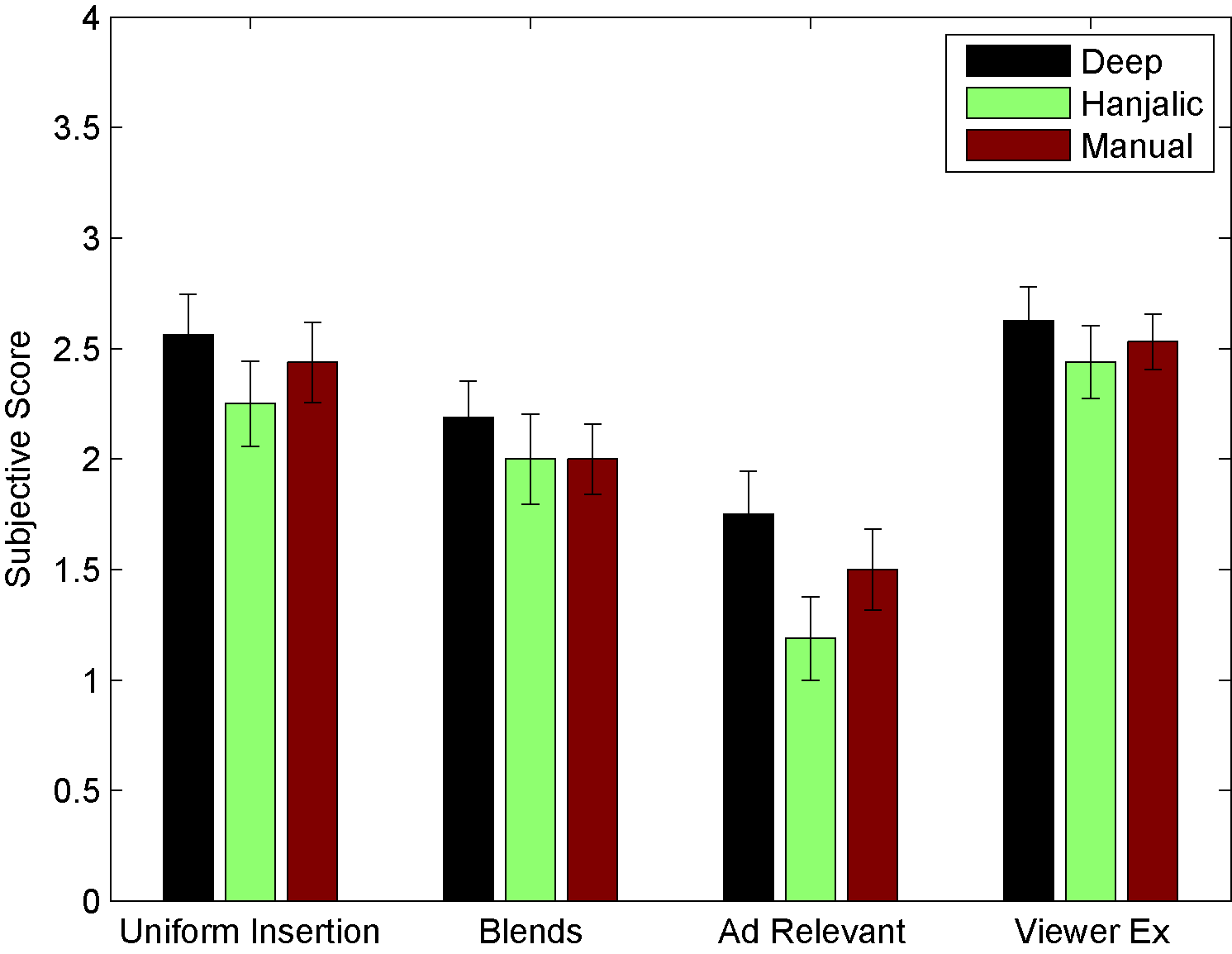}}
\centerline{\textbf{Immediate Recall}\hspace{0.2\linewidth}\textbf{Day-after recall}\hspace{0.23\linewidth}\textbf{User Experience}} \vspace{-.2cm}
\caption{\label{US_results} Summary of user study results in terms of recall and user experience-related measures.}
\vspace{-.2cm}
\end{figure*}

We evaluate the effectiveness of affective scores obtained from (a) our AAN-fc7 features, (b) Han features~\cite{Hanjalic2005} employed in CAVVA~\cite{cavva}, and (c) human assessments, for modeling contextual relevance based on user recall and experience responses.

Fig.~\ref{US_results} depicts user recall and experience-related measures as obtained with the three affect measurement approaches. Focusing on recall, although ad insertion via Han method~\cite{Hanjalic2005} results in higher immediate and day-after ad recall, lower incorrect recall and forgottenness ($p < 0.05$ in all cases), video programs generated with AAN-based affective scores are found to maximize user experience ($p < 0.05$ for insertion uniformity and ad relevance, and $p < 0.1$ for non-disruptive ad insertion). The significance of these effects was assessed by comparing the proportion mean  distributions for each question and method, via independent $t$-tests. Ads placed via the AAN model and correctly recalled by users were `well inserted' (difference with respect to manual scores significant at $p < 0.05$) based on responses compiled immediately upon viewing. 

Notably, ad insertions via AAN-based affective scores were opined to be (i) `uniformly distributed' across the streamed video, and  (ii) 'most relevant' in terms of emotional context with the video (Fig.\ref{US_results}). These observations imply that our AAN model is more accurate than Han in capturing the mood of the ads and video scenes. In contrast, while the Han method achieves the best recall from viewers, it also scores the least with respect to insertion-point distribution and relevance, implying an adverse impact on viewing experience. 


To examine \textbf{\textit{how affective attributes influenced ad recall}}, we correlated the ad recall measures with their (manually assigned) mean val, asl ratings. A meaningful relationship was noted between \textit{estimated ad valence} and the \textit{forgottenness rate} (Pearson $\rho = 0.45, p < 0.05$), indicating that positive val ads tend to be forgotten more easily. This observation agrees with the prior findings of Rimmele \etal~\cite{Rimmele}, who discovered that recall performance was maximum for negative valence images in a memory study. 

Surprisingly, ad insertions based on manual affective scores resulted in lowest recall and highest forgottenness among the three methods, while performing second best with respect to experience measures. This can be partly attributed to the higher val ratings observed for the selected ads based on manual scores ($\mu_{val} = 0.6$ for 12 unique inserted ads) as compared to ads selected based on AAN-based val estimates ($\mu_{val} = 0.3$ over 11 unique inserted ads). Ads selected based on Han val estimates had the lowest mean val ($\mu_{val} = 0.23$ over 12 unique inserted ads), suggesting that more low val ads were selected via the Han approach, resulting in least forgottenness. Nevertheless, viewers forgot nearly half the ads immediately and most ads a day later with all the considered methods. This reveals the scope for improving the ad-placement strategy by placing specific emphasis on ad retention.

To examine the \textbf{\textit{relationship between affective scores}} estimated by the three methods \textbf{\textit{and the inserted ads}}, we first examined if there was any relationship between manual ratings and computationally estimated scores. We found a significant correlation between Han-predicted and manual asl scores (Pearson $\rho = 0.4, p < 0.05$), but only a weakly significant correlation between manual ratings and AAN-based asl scores (Pearson $\rho = 0.24, p = 0.09$) on considering the 24 program video scenes and 28 ads (52 scores in total). Conversely, manual val ratings correlated significantly with our AAN model (Pearson $\rho = 0.45, p < 0.001$), but only weakly with Han estimates (Pearson $\rho = 0.25, p = 0.08$). 

The CAVVA optimization framework~\cite{cavva} has two components-- one for selection of ad insertion points into the program video, and another for selecting the set of ads to be inserted. Asl scores only play a role in the choice of insertion points, whereas valence scores influence both components. Our results suggest that accurate val prediction, as accomplished by our AAN model, plays a critical role in enhancing the subjective user experience. Although this improved experience seems to come at the expense of ad recall, we note that the Han method results in a disruptive experience despite high recall, and hence solely emphasizing on recall may not necessarily lead to the optimal ad placement strategy. 

\section{Conclusions and Future Work}\label{CFW}
This work discusses affect prediction from ads, and the utility of better ad affect estimation is demonstrated via a computational advertising application. A curative set of 100 diverse ads is compiled based on expert consensus, and its effectiveness as an affective dataset is examined based on ratings acquired from 14 raters. Dataset suitability is confirmed by (1) excellent agreement between the expert and annotator groups, and (2) uniform distribution of the asl and val ratings with minimal correlation between them.   

AAN-based audio-visual features are then proposed for encoding ad affect, and are found to significantly outperform features proposed by Hanjalic and Xu~\cite{Hanjalic2005} for val recognition. Best results with the AAN features are achieved with the MTL classifier, which effectively exploits the underlying audio-visual similarities among emotionally homogeneous ads. Finally, a study involving 17 users confirms that better modeling of ad emotions facilitates insertion of contextually relevant ads onto a streamed video. Specifically, the proposed AAN model is able to estimate ad valence better than the baseline~\cite{Hanjalic2005}, resulting in enhanced viewing experience. 

Future work will focus on the design of other informative (\eg, recurrent neural network-based) multimedia features for modeling affect. Another interesting line of inquiry is affect prediction via user physiological measurements similar to~\cite{Koelstra,decaf}. Finally, efficient approaches need to be designed for ad insertion within streamed video, so as to maximize both ad recall and viewing experience.

\begin{acks}
This research is supported by the National Research
Foundation, Prime Ministers Office, Singapore under its
International Research Centre in Singapore Funding Initiative.
\end{acks}

\bibliographystyle{ACM-Reference-Format}
\balance
\bibliography{affect_ads} 

\end{document}